\documentclass[10pt, conference]{IEEEtran}
\IEEEoverridecommandlockouts

\usepackage{cite}
\usepackage{amsmath,amssymb,amsfonts}
\usepackage{algorithmic}
\usepackage{graphicx}
\usepackage{textcomp}
\usepackage{todonotes}
\def\BibTeX{{\rm B\kern-.05em{\sc i\kern-.025em b}\kern-.08em
    T\kern-.1667em\lower.7ex\hbox{E}\kern-.125emX}}
\usepackage{listings}
\usepackage[framemethod=tikz]{mdframed}
\usepackage{color}
\usepackage{soul}
\usepackage{url}

\definecolor{lightgray}{rgb}{.9,.9,.9}
\definecolor{darkgray}{rgb}{.4,.4,.4}
\definecolor{purple}{rgb}{0.65, 0.12, 0.82}
\begin{document}

\title{6DVF: Data Visualisation Framework for mHealth Apps}
% \todo{give your framework a name...for example 6DVF...which stands for 6Dimensions  Visualisatin Framework}
% \todo{ if you mention mHealth, you need to explain in the paper why mheallth apps? and more importantly how mhealth viz is different from other types of apps-first para in the intro--done}

\author{\IEEEauthorblockN{Yasmeen Anjeer Alshehhi}
\IEEEauthorblockA{\textit{School of Inforamtion Technology} \\
\textit{Deakin University}\\
Burwood 3125, Australia \\
yanjeeralshehhi@deakin.edu.au}
\and
\IEEEauthorblockN{Khlood Ahmad }
\IEEEauthorblockA{\textit{School of Inforamtion Technology} \\
\textit{Deakin University}\\
Burwood 3125, Australia \\
ahmadkhl@deakin.edu.au}
\and
\IEEEauthorblockN{Mohamed Abdelrazek}
\IEEEauthorblockA{\textit{School of Inforamtion Technology} \\
\textit{Deakin University}\\
Burwood 3125, Australia \\
mohamed.abdelrazek@deakin.edu.au}
\and
\IEEEauthorblockN{Alessio Bonti}
\IEEEauthorblockA{\textit{School of Inforamtion Technology} \\
\textit{Deakin University}\\
Burwood 3125, Australia \\
a.bonti@deakin.edu.au}
}

\maketitle

%  overview of the topic 
% highlighting the importance of the topic, and/or
% Making general statements about the topic, and/or
% Presenting an overview on current research on the subject.

%  opposing an existing assumption, and/or
% Revealing a gap in existing research, and/or
% Formulating a research question or problem, and/or
% Continuing a disciplinary tradition.

% stating the intent of your study,
% Outlining the key characteristics of your study,
% Describing important results, and
% Giving a brief overview of the structure of the paper.
% \todo{mention a couple of existing frameworks--done}.
% \todo{ I did not understand this statement!!-done}. 
\begin{abstract}
The widespread of data visualisation tools on smartphones has provided end users an easy way to track their health data, leading designers to put more effort into delivering suitable visualisations.
Both academia and industry have developed several frameworks to guide the creation of informative and well-designed charts, such as the visualisation and design framework and Google Material Design. 
% Despite the  focus on design and chart types within existing frameworks, our study suggests the necessity for incorporating additional components in developing data visualisations. 
Despite the typical focus on design and chart types in these existing frameworks, our study highlights the need to incorporate additional components when developing data visualisations. The needs of non-expert users, the nature of the data being represented, and the mobile environment are often not prioritised in these frameworks, leading to visualisations that do not meet user needs and expectations. % The needs of non-expert users and the nature of the data being represented are often not the primary focus of these frameworks, resulting in visualisations that are not aligned with user needs and expectations.
To address these issues, we propose our Six-Dimensions Data Visualisation Framework (6DVF) to assist in the design and evaluation of visualisations on mobile devices. Finally, we present our initial findings from a designer evaluation experiment.
\end{abstract}


\section {Introduction} \label{intro}

% \textbf{Brief about data visualisation for health tracking :}
 Data visualisation is a crucial aspect of mHealth apps as this enables users to communicate effectively, make decisions, and identify trends using their collected health data. The ability to easily understand and make sense of this data is a key feature in mHealth apps.
 By displaying data over time, users can monitor their progress, detect patterns and trends, and gain valuable insights to improve their health outcomes.
 In light of the growing number of mobile health-tracking apps and the diverse user base, there is a need for a well-designed data visualisation framework that prioritises user experience (UX) and provides accurate and consistent visualisations on mobile devices. A well-designed data visualisation framework can enhance user engagement, provide actionable insights, and help users make informed decisions about their health.

Existing research, including Meyer et al. \cite{Meyer_2015}, Kelleher et al. \cite{kelleher2011ten}, Cuttone et al. \cite{Cuttone_2014}, and IBM \cite{IBM_2021}, have provided general guidelines for data visualisation design. 
While these guidelines encompass best practices for chart design, such as selecting charts, defining visualisation goals, they primarily focus on desktop computers \cite{Lee_2018}, and do not take into account the characteristics and context of non-expert users (i.e., those with limited knowledge of data visualisations).
Some studies have specifically explored data visualisation for tablet devices, offering guidelines for navigating the Roambi app\footnote{Roambi is an app used for creating business reports, dashboards, data visualisations, and charts \cite{Roambi_analytics}}\cite{r3}. 
In the industry, leading organisations such as IBM and Google have integrated data visualisation guidelines (IBM Design Language and Google Material Design) into their user interface frameworks to support chart design on mobile devices. However, these frameworks address general design elements and chart interactivity but do not consider user needs and chart functionality.

 In previous work, Alshehhi et al. conducted a review of user feedback (app reviews) on the mHealth app \cite{Yasmeen_2022}, a user survey to understand needs and opportunities for further development \cite{Alshehhi_2022}. They discovered that there was dissatisfaction among end-users and challenges in using charts to track health data. These difficulties encompassed various aspects of the charts such as functionality, data displayed, styling, lack of adaptability for diverse user groups and the overall data visualisation interface.

 This paper introduces a new data visualisation framework (6DVF) that aims to address the limitations identified in previous research. The 6DVF includes six dimensions: (1) identifying the target audience; (2) chart functionality; (3) data representation; (4) visual design and interactivity; (5) target device; and (6) single/app visualisation interface. Additionally, a checklist has been developed to evaluate the framework's output. Experiments are being conducted to assess the framework's effectiveness, involving evaluations by designers and end-users. The designers are asked to create data visualisations with and without the framework, which are then evaluated with end-users to measure the impact to user experience (UX).

 The rest of this paper is structured as follows, section \ref{Related-work} elaborates on the gaps found in the current frameworks. Section \ref{framewrok-dev} explains the proposed framework's development process. Section \ref{framewrok-eval} details the evaluation plan to evaluate the framework. Section \ref{results} elaborates on the preliminary results. Finally, section \ref{Discuss} discusses the potential use of this framework and briefs the work presented in this paper.

\section{Related-work} 
\label{Related-work}

The academic field has produced numerous studies on developing data visualisation frameworks and guidelines. Some studies, such as Cuttone et al. \cite{Cuttone_2014}, have focused on reducing cognitive load when using data visualisation to explore personal information. They focused on the presented data, the suitable chart representing time series and comparison data, and included simple interactivity. However, the study did not include non-expert users, which raises concerns about users' familiarity with the presented data and charts. 
Meyer's work \cite{Meyer_2015} offers a comprehensive set of guidelines that consider the targeted users, tasks, and data when designing data visualisations. However, it remains to be seen if these guidelines can be applied to mobile device scenarios, raising concerns about the scalability of charts due to screen size and input methods. 
Lee et al. \cite{Kandogan_2016} conducted a grounded theory study to investigate the current efforts related to data visualisation guidelines. They found that current guidelines focus on users' ability to understand charts, users with colour blindness disabilities, visualisation components, tasks, data, and devices. However, mobile devices were not included in these guidelines, as the study shows that inputs were made either using a mouse or a keyboard. Additionally, data visualisation styling was separate from the aspects covered in the collected guidelines. Grainger et al. \cite{Grainger_2016} focused on exploring non-expert users' characterisation, data visualisation design, and styling, which supports the idea of the need to understand users.

The industrial sector also includes two main frameworks for developing data visualisation for mobile devices. 
Google Material design \cite{GoogleMaterial} is a design language that offers guidelines for creating user interfaces. It includes data visualisation design as one important aspect and focuses on the chart type, style, and behaviour. 
Apple's Human interface guidelines \cite{HIG_Apple} provides general advice on - (1) delivering the best practices related to data visualisation design, such as chart presentation; (2) the best practices for data visualisation; and (3) enhancing chart accessibility. However, there are limitations to the interaction with these charts and platforms.
 
Despite significant efforts in developing guidelines for data visualisation, the rapid advancement of technology presents challenges for them to continue being a reliable source of guidance \cite{r2}. When designing data visualisations, it is also important to consider the context of use, such as: 

\begin{itemize}
    \item \textbf{User needs}: The context of mHealth tracking includes plotting real-time data, periodically tracking data (daily, weekly, monthly, and yearly), and finding correlations between different data types. Therefore, identifying users' needs while exploring their charted data is essential. Additional needs include chart usability, accessibility for users with disabilities and older adults, and better understanding of charted data.
    \item \textbf{Chart interactivity}: Allows users to explore and interact with their tracked data, which in turn aids in better understanding of their data.
    \item \textbf{Mobile devices}: Allows users to access their charts anytime. Therefore, it is crucial to consider the chart scalability with the screen size and the touch-based input modality \cite{Katy_2019}.
\end{itemize}

In summary, considering the context of use, users' needs and the specific characteristics of mobile devices when designing data visualisations for mHealth apps can help ensure that visualisations are legible, usable, and efficient on these devices, which in turn can improve the UX and make the data more accessible to users.

% Several attempts have been made to build frameworks for data visualisation design. Interactivity, cite{wong_18}, \cite{NW_2021}, 

%  ,

% We found one framework that focused on interaction modality and touch surface \cite{r3}. Further frameworks targeted non-scientific audiences ,, and another framework focused on the effectiveness of data visualisation \cite{Zhu_2007}. 

% Finally, the above studies identified general recommendations and guidelines for designing data visualisation. However, this paper acknowledges the need to build flexible frameworks for mHealth data visualisation designers/app developers to be used in multiple mHealth tracking scenarios.

\section{Framework Development} 
\label{framewrok-dev}
% \todo{you need to add few sentences here explaining what this section covers}
This section presents the details of the components that make up our 6DVF and how they work together. It also includes a checklist for developing and maintaining the framework.
% \todo{fig1 does nt make sense...
% i think it is user --> functionality --> data --> device --> design sys --> data visualisations}

% \todo{I would move this to the intro...see my cocmments related to including your work pararaph--done}
% \begin{figure}
%   \centering
%   \includegraphics[width=1\linewidth] {Images_ch7/steps}
%   \caption{Framework architecture}
%   \label{fig:steps}
%   % \vspace{-18pt}
% \end{figure}

% \todo{mention 6Dimensions model/process}

\subsection {The 6 Dimensions}
% As we discussed, 6DVF
%the proposed framework for data visualisation design in mHealth tracking applications
As we discussed, 6DVF builds on existing guidelines and includes additional considerations for non-expert users and the context of mhealth tracking. We incorporated six dimensions to enable the delivery of best practices in data visualisation. These six dimensions are divided into two main parts: 1) Empathize and Needs, which addresses user understanding and requirement identification; and 2) Ideate, which presents the necessary components for constructing a data visualisation interface using a design system approach. Each of these dimensions includes an I/O process.

% (figure \ref{fig:frameworkflow}).

% \todo{if we have space, you should highlight the inputs and outputs of each step or dimension -- remember the slides you prepared for the participants.. so it becomes more concrete. At the mooment it sounds like a thought without concrete example}

\subsubsection{Empathize \& Needs}
\begin{figure}
  \centering
  \includegraphics[width=1\linewidth] {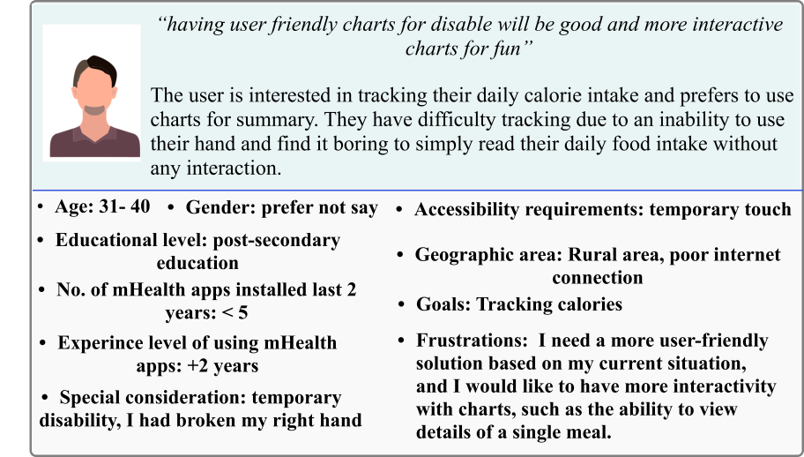}
  \caption{A sample of the collected persona}
  \label{fig:persona}
  % \vspace{-18pt}
\end{figure}

This section aims to understand users, define their needs, and identify mobile device features. The following components are considered:

\textit{Non-expert user characterisation:} includes gaining a deeper understanding of the target audience through quantitative research methods such as market research and user surveys. This information can create user personas and make data visualisation more accessible to a broader range of users. 
The output of this process is a set of personas. Figure \ref{fig:persona} shows a sample of our persona collection.

\textit{Users' needs and pain points:} defines the problems, emotions, and experiences users have when using data visualisation. This information helps create an ideal user journey that outlines users' actions, thoughts, pain points, and opportunities when visualising data.
% The user journey, as depicted in Figure \ref{fig:userjourney}, is based on the data collected.

% \begin{figure}
%   \centering
%   \includegraphics[width=1\linewidth] {Images_ch7/userjourney}
%   \caption{User journey}
%   \label{fig:userjoureny}
%   % \vspace{-18pt}
% \end{figure}

\textit{Target platforms:} considers mobile devices, including tablets, smartphones, and wearable devices. This includes taking into account features such as screen size and interactive capabilities to ensure optimal visualisation presentation and address accessibility needs. A literature survey or market research would be an ideal method to understand the features of the current devices.

\subsubsection{Ideate}

The following components are considered in this section:

\textit{Data:} identifies the data captured to determine the suitable charts and patterns to make sense of the collected data. This enables the creation of different charts using chart tasks such as finding anomalies, correlations, and periodic summaries. The output of this process is mapping the data collected with suitable charts and a list of sketched data visualisations. 

\textit{Design system:} identifies two components which are: \textit{Look and feel:} represents the chart design and appearance, including a colour palette, chart layout, font sizes, labelling and types, and accessibility options symbols. \textit{Interactivity:} addresses user interaction with the charts and the data used to build the charts. It also includes the interaction mechanism limited to tap, pinch, swap, gestures and voice notes.

\textit{App visualisations:} involves the final appearance of the data visualisation and incorporates all the elements identified in the previous steps. To ensure the best possible outcome, designers need to repeat this step for every visualisation.

% is an itertiavie step in which desingers need to repeat it for each visualisarion have been idetified earlier. It represents the final look of the data visualisation, which incorporates the above three components. 

Overall, %this framework
6DVF provides a comprehensive approach to designing and developing compelling and informative data visualisations for non-expert users in mHealth tracking applications by considering the user's perspective, needs and device features

\subsection{Framework Checklist}

We prepared a checklist to support designers in evaluating the developed data visualisation. The list provides an efficient pace of evaluation, as stated in \cite{Sawicki_2022}. It includes the framework components we mentioned in Section \ref{framewrok-dev} to be validated using the Consistent, Complete, and Correct (3Cs) criteria, as shown in Table \ref{tab:FWchecklist}. The 3Cs criteria are known to be used for software requirement specifications validation, as stated in \cite{Kamalrudin_2015}. In the data visualisation context, the 3Cs standards enable designers to deliver reliable charts that meet users' needs.

\begin{itemize}
    \item Complete: is to ensure that all components of a given dimension have been built to meet users' needs.
    \item Consistent: refers to the consistency of all components of the data visualisation interface.
    \item Correct: refers to the error-free in all the components captured or produced in each dimension.
\end{itemize}

\begin{table*}
\caption{Framework checklist}
\label{tab:FWchecklist}
\begin{center}
  \begin{tabular}{ |p{4cm}|p{4cm}|p{4cm}|p{4cm}|  }
    \hline
    \multicolumn{4}{|c|}{Visualisation Criteria} \\
    \hline
    Visualisation Dimension& Complete & Consistent & Correct  \\
    \hline
    Targeted Audience   & Do we have a complete list of the target audience's characteristics (target users) for the data visualisation?
    &Is there any conflict between the target audience (users) of the data visualisation
    &   Do we have the correct audience?
    \\
    Functional Requirement&   Do we have a complete list of data visualisation functional requirements
    & Is there any conflict or inconsistencies between the functional requirements
    &Are these functional requirements correct
    \\
    Data &Do we have all the data required to achieve the intended data visualisation
     & Is data consistent and can be linked – i.e. same granularity in terms of special, temporal, units of measure
    &  Do we have the correct data for the visualisation
    \\
    Design system (Look and feel \& interactivity)   &Do we have all the look and feel (non-functional) requirements for the data visualisation
     & Do we have a consistent look and feel throughout all the visualisation in the app
    &  Do we use the correct look and feel in all app visualisation
    \\
    Targeted device&   Do we consider all the device and platform capabilities, limitations, and compatibility to present the required data visualisation
    & Are all the visualisation consistent with the limitations and capabilities of the underlying platform
    &Is the visualisation style consistent with the user profile / data / functional requirements
    \\
     Single/App visualisation & Does a visualisation cover all (functional and non-functional) the user requirements meant for this specific visualisation
     & Is the visualisation style consistent with the user profile / data / functional requirements
     &Are these the correct data visualisation that the users need?
    \\

     \hline

    \end{tabular}
    \end{center}
\end{table*}

\section{Evaluation} 
\label{framewrok-eval}

% \todo{thinking about it again, you could include a usage example that you show how youu uused uur framewrk to builld data viisualuusations for the diet app yoou built--
% Yasmeen : So it is either shwoing the results or the usage exmaple ?
% }

\begin{figure}
  \centering
  \includegraphics[width=.80\linewidth] {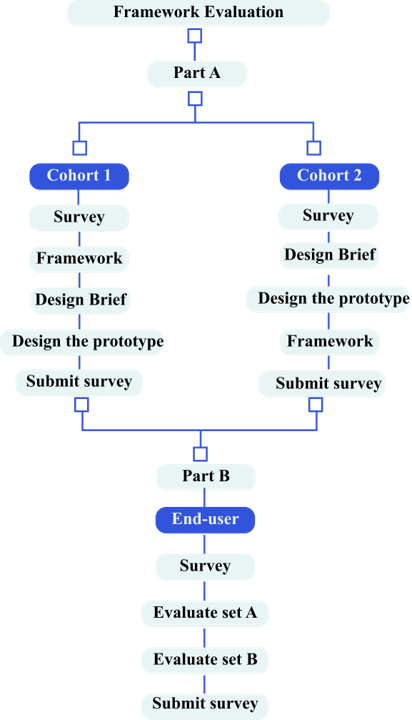}
  \caption{The flow of the evaluation plan}
  \label{fig:evaluationplan}
  % \vspace{-18pt}
\end{figure}

The evaluation plan for this study involves both user testing and A/B testing approaches. The user testing approach is being used to evaluate the framework from the perspectives of designers and app developers. In contrast, the A/B testing approach examines the data visualisations created using the framework from end-users' perspectives. The experiment is divided into two main parts, as shown in figure \ref{fig:evaluationplan}.

\subsection{Part A: Designer Study} \label{partA}
In this part, we implemented the user testing approach to evaluate the framework in real-life scenarios. We aim to recruit app developers, data visualisation designers, and UX designers to create a meal-tracking app prototype. In this experiment, participants are divided into two groups:

\begin{itemize}
    \item Group 1 (Cohort 1): Participants are introduced to the framework and presented with a case study of a mobile data visualisation scenario. They are then asked to create a set of data visualisations and complete a survey.
    \item Group 2 (Cohort 2): Participants are not made familiar with the framework, but are also presented with the same case study. They are then asked to create a set of data visualisations. After completing this task, we introduce them to the framework and ask them to complete a survey.
\end{itemize}

\subsection{Part B: End-user Study}
The end-users will receive a link to a survey in which they will evaluate sets A and B of the meal-tracking prototypes designed by the participants in \ref{partA}. Each set will randomly include three different prototypes. This will allow us to compare the effectiveness and usability of the prototypes created using the framework to those created without it, providing insights into the framework's strengths and weaknesses from the end user's perspective.

\section{Preliminary results and discussion}
\label{results}

% We have initiated the experiment and have received four responses thus far, two from each cohort.
% Both designers from cohort 1 reported that the framework was easy to use and helped them design their meal tracker app prototype. They also expressed confidence in using the framework. In addition, the designers provided neutral and positive feedback on their likelihood of recommending it to other designers and their frequency of usage. With respect to cohort 2 designers, both designers believed that using the framework would enable them to deliver better chart designs that would suit user needs. However, one designer noted that the framework might introduce some limitations regarding flexibility related to the data visualisation context.
% Both charts in figure \ref{fig:outcomesmaple} show the relationship between meals and the money spent. The chart on the left was built using the 6DVF framework, while the chart on the right was built without following the framework. The left chart includes different accessibility options, such as chart titles and text explanations, making it more accessible to users with disabilities. It also includes simple interactivity, allowing users to edit their charts. In contrast, the right chart does not include accessibility options or interactivity. Overall, the left chart provides a more inclusive and user-friendly experience.

% In our study, we sought to assess the feasibility and usefulness of the 6DVF framework for data visualisation. 
We have initiated our evaluation and have received six responses thus far, three from each cohort. The experiment results revealed a positive impact of the framework on the designers' ability to create accessible and user-friendly charts. Designers from cohort 1 reported that the framework was easy to use and helped them in designing their meal tracker app prototype. 
% This is a promising outcome, as the ease of use and accessibility of the framework can help designers to create more user-centred designs.
The feedback from cohort 2 designers also highlighted the potential of the 6DVF framework to improve chart design practices. The designers expressed confidence in using the framework and noted its potential to deliver better chart designs that would meet user needs. However, one designer expressed some concerns about the framework's limitations in terms of flexibility. This is an important consideration that should be addressed in future research to ensure that the framework is useful and relevant to a wider range of data visualisation contexts.

\begin{figure}
  \centering
  \includegraphics[width=.80\linewidth] {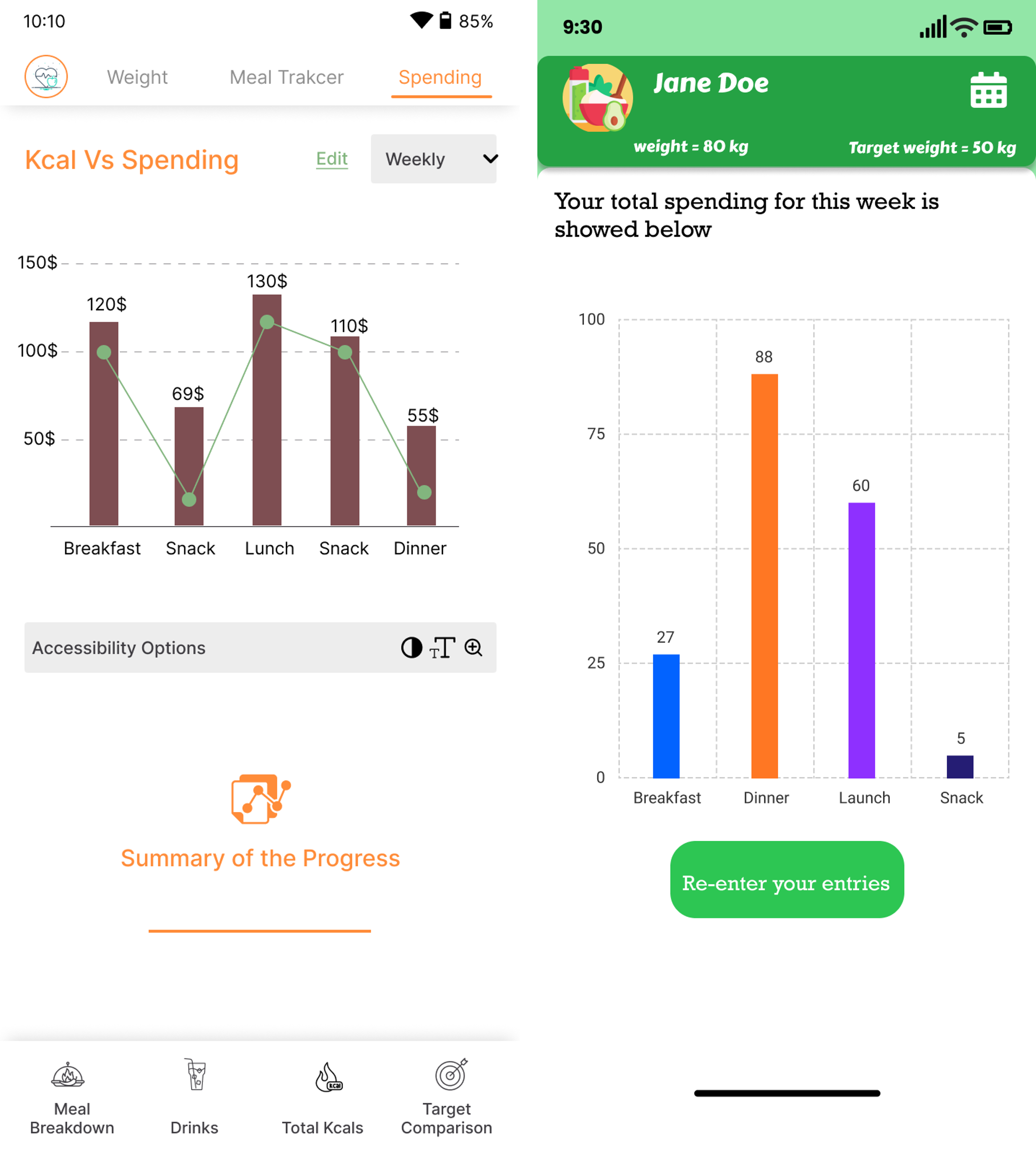}
  \caption{Outcome samples - visualisation using our framework (left) and without (right).}
  \label{fig:outcomesmaple}
  % \vspace{-18pt}
\end{figure}

The comparison of the two charts in figure \ref{fig:outcomesmaple} demonstrates the impact of the 6DVF framework on chart design. The chart built using the framework includes various accessibility options, such as chart titles and text explanations, making it more accessible to users with disabilities. This is a crucial aspect of data visualisation design, ensuring that charts are inclusive and usable for everyone. The chart also includes simple interactivity, allowing users to edit their charts. In contrast, the chart built without following the framework does not include accessibility options or interactivity, providing a less user-friendly experience.

In conclusion, our experiment highlights the potential of the 6DVF framework to improve data visualisation design practices and enhance user experience on mobile devices. While the results are preliminary, they provide a foundation for future research and development of the framework. The positive feedback from designers suggests that the framework has the potential to be widely adopted and help designers create accessible and user-centred charts.

\section{Conclusion} 
\label{Discuss}
% A significant challenge is the need for a comprehensive understanding of end-users.

Despite the availability of various frameworks and guidelines in academic and industrial sectors, there needs to be more emphasis on end-users needs.
To address these gaps, we explored users' perspectives, challenges, and expectations regarding charts. 
Our 6DVF approach is built on the findings of these studies and serves as a starting point for both users to adapt to charts and designers to implement best practices. In addition, it considers the unique features, accessibility, and scalability of mobile devices.
Our framework represents the first step towards personalised data visualisation, enabling users to customise chart colours, layouts, interactivity, and annotations.

\bibliographystyle{IEEEtran}
\bibliography{references}

\end{document}